# Determination and Spectroscopy of Quantum Yields in Bio/Chemiluminescence via Novel Light-Collection-Efficiency Calibration: Reexamination of The Aqueous Luminol Chemiluminescence Standard

Yoriko Ando[1,2*], Kazuki Niwa[3], Nobuyuki Yamada[4], Tsutomu Irie[4], Toshiteru Enomoto[4], Hidehiro Kubota[2,4], Yoshihiro Ohmiya[3], Hidefumi Akiyama[1,2]

[1] Institute for Solid State Physics, University of Tokyo, Kashiwa, Chiba, Japan

[2] CREST JST, Kawaguchi, Saitama, Japan

[3] AIST, Ikeda, Osaka, Japan

[4] ATTO Corp., Bunkyo, Tokyo, Japan

Institute for Solid State Physics, University of Tokyo

A273, 5-1-5 Kashiwanoha, Kashiwa, Chiba, 277-8581, Japan

**Phone**: +81-4-7136-3387

**Fax:** +81-4-7136-3388

**e-mail**: yori@issp.u-tokyo.ac.jp

**Abbreviations:** no abbreviations

[*] To whom correspondence should be addressed at Institute for Solid State Physics, University of Tokyo, A273, 5-1-5 Kashiwanoha, Kashiwa, Chiba, 277-8581, Japan. Phone: +81-4-7136-3387, Fax: +81-4-7136-3388, e-mail: yori@issp.u-tokyo.ac.jp

**ABSTRACT**

We have developed a luminescence-measurement system for liquid bio/chemiluminescence that can obtain quantitative luminescence spectra as the absolute total number of luminescence photons at each wavelength or photon energy and quantum yields. Calibration of light-collection efficiency in the system is performed with a reference double-plate cell. This method is applicable to sample cells of any kind suitable for measurement, which is a great advantage over previous techniques in practical experiments. Using this system, the quantum yield of aqueous luminol chemiluminescence was obtained as 1.23 ± 0.20%, which is in good agreement with previously reported values.

## INTRODUCTION

Luminescence intensities and spectra are usually measured in arbitrary units or relative light units (RLU) because the determination of absolute luminescence intensity spreading over all directions needs painstaking calibrations for the geometrical light-collection efficiency and wavelength-dependent absolute sensitivity of a detector, which are not very easy in practice. However, presentations of quantitative luminescence intensities and spectra scaled in absolute light units allow direct comparisons among independently measured data, which may provide new insight from the rich accumulation of data.

Quantum yield in bio/chemiluminescence is defined as the probability of photon emission via the reaction of a single substrate molecule and is a key quantity to characterize the reaction, understand its mechanisms microscopically, and also develop its applications. Experimental determination of the quantum yield is achieved by dividing the absolute total number of luminescence photons, or the integrated quanta of luminescence, by the number of consumed substrate molecules. Quantum yields of popular bio/chemiluminescence reaction systems such as firefly (1), cypridina (2), aequorin (3-5), and luminol (6) were mostly reported in 1960 - 1970. Luminol chemiluminescence was later reexamined by other elaborate methods (7-13) and is now often used as one of secondary light standards (6, 14) in luminescence measurements. In most of these experiments, however, the absolute sensitivities of detectors were calibrated individually for their respective luminescence colors or spectra by multi-step calibrations. Furthermore, to estimate geometrical light collection efficiencies assuming a point-source approximation, sample cells and measurement systems with special geometries and large sizes were selected. To overcome such inconveniences, we need to devise more useful methods of quantitative luminescence measurement and calibration for any luminescence colors and sample cells.

For this purpose, we have developed a system for quantitatively measuring luminescence yield spectra. The luminescence yield spectrum is defined as a quantitative luminescence spectrum, where the absolute number of photons emitted as luminescence in all directions is plotted as a function of wavelength or photon energy. To calibrate light collection efficiencies for sample cells of any kinds, we devised a reference double-plate cell and accurately determined the efficiencies for various sample cells by comparison. Additional calibration of the wavelength-dependent absolute sensitivity of the detector-spectrometer system enabled measurements of luminescence yield spectra.

The integrated area of the luminescence yield spectrum gives the absolute integrated number of luminescence photons, or the integrated quanta of luminescence, and its division by the number of consumed substrate molecules gives the quantum yield. Note that the above-defined luminescence yield spectrum divided by the number of substrate molecules, or the luminescence spectrum plotted in the absolute unit of the number of photons per substrate molecule as a function of wavelength or photon energy, is nothing but the spectrum of the quantum yield, because the integrated area of this luminescence spectrum is equal to the quantum yield. It is probably the best way to quantitatively present a luminescence spectrum for direct comparison with other bio/chemiluminescence systems.

In this paper, we describe our quantitative bio/chemiluminescence measurement system and calibration methods. In particular, a new method of light-collection-efficiency calibration using a reference double-plate cell is described in detail, and calibrated values for several sample cells are shown. Using this system, we measured absolute luminescence yield spectra and quantum yields of aqueous luminol chemiluminescence catalyzed by horseradish peroxidase to check consistency with other reports (6-13). We obtained the quantum yield of 1.23 ± 0.20%, which is in good agreement with the previously reported values. We discuss the use of luminol chemiluminescence as a light standard and the advantages of our measurement system as well as calibrations.

## MATERIALS AND METHODS

*Luminescence measurement system.* A schematic drawing of the experimental setup for the luminescence yield measurement of bio/chemi-luminescence is shown in Figure 1. We prepared two solutions: a substrate solution and a trigger solution. The substrate solution was put in a sample cell, to which the trigger solution was added quickly with a micropipette to initiate reaction. Any type of sample cell with a window size below about 6 mm x 6 mm can be used, and we have tested several kinds: a homemade acrylate tube cell and commercially available transparent, black, and white 384-well microplate cells, as illustrated in Figure 2. When we used the acrylate tube cell, luminescence was measured from the bottom window of the cell. For the 384-well microplate cells, it was measured from the top opening of the cells. The volume of the solution was typically 100 μl.

Luminescence from the cell was reflected by a mirror, collected by a lens with a focal length of 50 mm placed 87 mm away from the cell, and focused on the entrance slit of a 300 mm-focal-length spectrometer (SpectraPro 300i, Acton Research) with F: 4 optics. An iris with an opening of diameter $r$ = 20 mm was placed in front of the lens ($a$ = 73.0 mm away from the cell) to define a solid angle of luminescence collection, or a numerical aperture $NA = \sin\theta$, where, $\theta = \tan^{-1}(r/2a)$ is the maximum angle of light collection. We narrowed the entrance slit of the spectrometer typically to 0.25 mm and selected a diffraction grating, which caused the luminescence to be spectrally dispersed on the detector plane at the exit port of the spectrometer.

As a sensitive photodetector, we used a back-illuminated CCD camera (CCD-1100-PB, Princeton Instruments) cooled to -120 $^{o}$C with liquid nitrogen. A mechanical shutter placed in front of the CCD detector controlled exposure time. We started exposure just before we initiated the reaction by adding a trigger solution to the substrate solution. An exposure for 10 s and a read-out for 75 ms were alternately repeated until the reaction was completed.

To determine the absolute luminescence yield spectrum scaled by the number of photons for bio/chemiluminescence in a solution through the above measurement, two kinds of calibrations are needed: for light-collection efficiency $\eta_{\mathrm{cell}}$ with slit transmissivity $T$ and absolute sensitivity $S(\lambda)$ of the CCD-spectrometer system. On the basis of these calibrations, the obtained spectrum $C(\lambda)$ of bio/chemiluminescence in units of CCD counts/nm gave luminescence yield spectrum $N_\lambda$ or $N_E$ in units of nm$^{-1}$ or eV$^{-1}$, from which the quantum yield was determined.

*Calibration.* A) Light-collection efficiency. In the setup shown in Figure 1, the iris and the lens collect only a portion of the total luminescence. This fraction, which we call the light-collection efficiency, must be known in order to derive the luminescence yield. We developed a reference double-plate cell, which has a very simple structure and enable the collection efficiency to be obtained by simple geometrical calculation. Then we put a reference luminescent solution into the reference double-plate cell and a sample cell used for the luminescence measurement and experimentally compared these luminescence intensities to obtain the collection efficiency of the sample cell. The structure of the reference double-plate cell is illustrated in Figure 2. It consists of three layers of glass plates, where a 0.5-mm-thick gap with a 30 mm x 50 mm rectangular area is formed between two 0.9-mm-thick glass plates. We filled the gap with a reference luminescent solution. For this calibration, a reference luminescent solution with constant luminescence intensity during the calibration is convenient. So we used luminescent solution, PLT7.5 (TOYO B-Net, Tokyo, Japan), which is a firefly bioluminescence-solution kit having very stable luminescence with a very long half-life of 7.5 hours.

To limit and define the measured volume of the luminescence solution, a black mask with a circular hole 4.0 mm in diameter in the center was put on the double-plate cell, and luminescence emerging from this hole was collected by the iris-lens system, as shown in Figures 2 and 3. (While a mirror is not illustrated in Figure 3 for simplicity, this calibration was done with the setup shown in Figure 1.) This let us measure luminescence from only a volume $(4.0/2)^2 \pi$ x 0.5 $mm^3$, or 6.3 µl. Partial reflectance of light at the air-glass and glass-solution interfaces did not cause loss, because the same effects occurred at the upper and lower interfaces and cancelled each other out. We assumed that the same luminescence intensities were emitted upward and downward from the double-plate cell. Provided that the area of the hole of the black mask was much smaller than the whole area of the double-plate cell, the reflection or scattering at the edge of the double-plate cell was negligible.

The collection efficiency is easily calculated for this double-plate cell, as shown below. The maximum angle $\theta = \tan^{-1}(r/2a)$ of light collection is limited by the iris and the lens with aperture of diameter $r$ at distance $a$, which defines an external numerical aperture $NA = \sin\theta$, shown as Figure 3. The external maximum angle $\theta$ is related to the internal maximum angle $\theta_i$ via $\sin\theta_i = \sin\theta / n_i$, where $n_i$ is the refractive index of a solution. Thus, the internal numerical aperture $NA_i$ is given by $NA_i = NA / n_i$. The collection efficiency $\eta_{plate}$ is given by the ratio of the solid angle corresponding to $NA_i$ versus 4π, as

$$\eta_{plate} = \frac{1-\sqrt{1-(NA/n_i)^2}}{2}. \qquad .1$$

For our case of $r$ = 20.0 mm, $a$ = 73.0 mm, and $n_i$ = 1.34, $NA$ is 0.136, and the collection efficiency was $\eta_{plate}$ = 0.262% for the double-plate cell.

We then measured the luminescence of the same solution using the sample cell. By comparing measured luminescence intensities per unit volume of solution, we obtained the ratio of collection efficiencies with the sample cell and the reference double-plate cell, $\eta_{cell} / \eta_{plate}$. As the collection efficiency $\eta_{plate}$ of the reference double-plate cell had already been obtained by the above geometrical calculation, the measured ratio gave the collection efficiency $\eta_{cell}$ for the sample cell.

We tested several kinds of sample cells: a homemade acrylate tube cell and commercially available transparent, black, and white 384-well microplate cells, as illustrated in Figure 2. The acrylate tube cell was made of a transparent acrylate tube with 4-mm inner diameter, 6-mm outer diameter, and 20-mm length glued to a 0.5-mm-thick transparent acrylate plate. Luminescence was measured via the bottom. The 384-well microplate had 384 cells with the dimensions of 3.5 mm x 3.5 mm x 10.4 mm. Solution with a volume of 100 µl almost completely filled the cell. Luminescence was measured via the 3.5 mm x 3.5 mm top openings.

The collection efficiency for these cells with 100-µl luminescent solutions calibrated in reference to a double-plate cell are listed in Table 1. Though the acrylate tube cell had a rather complicated structure and a light-piping effect, the measured collection efficiency 0.26 ± 0.01% with 100-µl luminescent solution was close to that of the reference double-plate cell. The measured efficiency of 0.18 ± 0.02% for the transparent 384-well microplate cell was slightly smaller than the values for the reference double-plate cell and the acrylate tube cell, possibly because of light collection from the top openings. The efficiency of 0.16 ± 0.02% for the black 384-well microplate cell was even smaller. This is because light reflected by cell walls and reaching the top surface with small incident angle could be collected with the transparent cell, while it was absorbed in the black cells.

The white 384-well microplate cell showed high collection efficiency of 1.7 ± 0.2%, which is about 10 times larger than that of the transparent and black cells. This is because the white cell had an effect like an integration sphere, so the luminescence light repeatedly experienced diffused reflection, or scattering, on the white cell walls until all the light came out of the solution from the top opening. The use of the white cell needs some care. One is issue that the most commercial white cells have wavelength-dependent reflectivity. Another is that they often shows long-lived phosphorescence by themselves in the near infrared wavelength region. Though the white cell has these problems, its high collection efficiency is a great advantage.

We also measured collection efficiencies for different volumes of the luminescent solution, as shown in Figure 4. Note that the collection efficiency was fairly flat for the acrylate tube cell, though it decreased with increasing volume of the solution. Other 384-well microplate cells showed strong volume dependence of $\eta_{\mathrm{cell}}$. In particular, the transparent and black 384-well microplate cells showed sensitive changes in $\eta_{\mathrm{cell}}$ near 100 μl, when the cell was filled and shape of the liquid surface changed from concave to convex. This is possibly related to the relatively large errors in the calibration of $\eta_{\mathrm{cell}}$ for those cells. On the basis of this calibration, for the luminol luminescence measurement, we selected and used the acrylate tube cell, which had the least volume dependence and fluctuation in the collection efficiency.

The entrance-slit width of the spectrometer was reduced typically to 0.25 mm, which passed only a small portion of luminescence. This transmissivity $T$ should also be calibrated. For this calibration, we switched the grating in the spectrometer to a mirror and measured images and intensities of the luminescence from the cell with and without the 0.25-mm-wide slit with the CCD camera. From the luminescence intensity ratio, the transmissivity $T$ was determined.

B) Sensitivity of CCD-spectrometer system. We calibrated the absolute sensitivity $S(\lambda)$ of the CCD-spectrometer system for visible wavelengths λ in two steps: 1) calibration of the absolute sensitivity at certain wavelengths and 2) calibration of relative sensitivity for all wavelengths. We first calibrated the absolute sensitivity $S_{\lambda 0}$ of the system with laser light. Using a calibrated power meter, we measured the absolute power of a light beam from a laser. We measured CCD counts for the correctly attenuated laser light to obtain the absolute sensitivity of the system in units of counts/photon. For the second step of calibrating the relative sensitivity of the system for all wavelengths, we used a tungsten lamp with an optical fiber output. We measured the spectrum of the output light in the wavelength region covering 350 - 800 nm, with a calibrated optical spectrum analyzer. Then we measured the output light from the tungsten lamp with the CCD-spectrometer system to obtain the sensitivity $\varepsilon(\lambda)$ in arbitrary units. Multiplying the absolute sensitivity $S_{\lambda 0}$ at a laser wavelength $\lambda_0$ and the relative sensitivity of the detection system $\varepsilon(\lambda) / \varepsilon_{\lambda 0}$, we obtained the absolute sensitivity $S(\lambda) = S_{\lambda 0}$ x $\varepsilon(\lambda) /\varepsilon_{\lambda 0}$.

*Data acquisition and conversion*. A spectrum's CCD counts per wavelength $C(\lambda)$ (counts/nm) was obtained as raw data of the luminescence measurement. We obtained a luminescence yield spectrum $N(\lambda)$ ($nm^{-1}$) by dividing $C(\lambda)$ by the geometrical collection efficiency $\eta_{cell}$ , the transmissivity $T$ at the spectrometer slit, and the absolute sensitivity of CCD-spectrometer

system $S(\lambda)$ as

$$N_\lambda = \frac{C(\lambda)}{\eta_{\text{cell}} T S(\lambda)}. \qquad (2)$$

The spectrum $N_\lambda$ can be converted into spectrum $N_E$ ($eV^{-1}$) as a function of photon energy $E$(eV) as

$$N_E = N_\lambda \times \frac{d\lambda}{dE}. \qquad (3)$$

A spectrum as a function of photon energy $E$ instead of wavelength λ is often preferred in the discussion of microscopic mechanisms and origins because it is suitable for spectral analysis assuming some transition energies and energy broadening inherent to possible excited-state molecules.

By integrating $N_\lambda$ or $N_E$ with λ or $E$, we obtained the total number of luminescence photons $N_{\text{total}}$ as

$$N_{\text{total}} = \int N_\lambda d\lambda = \int N_E dE. \qquad (4)$$

Quantum yield was obtained by dividing $N_{\text{total}}$ by the number of substrate molecules $M$ in the solution.

Note that the luminescence spectrum $N_E / M$ obtained by dividing the luminescence yield spectrum $N_E$ by the number $M$ of substrate molecules is a quantitative luminescence spectrum such that its integrated area is equal to the quantum yield. Therefore, it is nothing but the spectrum of the quantum yield. One can quantitatively compare spectra of quantum yields as well as quantum yield values among different bio/chemiluminescence systems under different conditions.

*Sources of uncertainty.* Estimated uncertainties are listed in Table 2. To calibrate the light-collection efficiency, we calculated $\eta_{plate}$ from $NA$ on the basis of measured geometrical lengths $r$ and $a$, where we estimated the uncertainty in $NA$ as 0.3%. Since $\eta_{plate}$ was approximately proportional to $NA^2$, the uncertainty introduced into $\eta_{plate}$ was 0.3 x 2%. We estimated an uncertainty of 10% in the relative efficiency measurements with the acrylate tube cell. The uncertainty in the slit transmissivity calibration was estimated to be 3.4% from empirical fluctuations of calibration values. Thus, the total uncertainty in the light-collection-efficiency calibration was estimated to be 10.6%. We also estimated the uncertainty in calibration of the CCD-spectrometer system as 12%, which is due to uncertainties of 9.4% in the absolute sensitivity $S_{\lambda 0}$ and of 7.1% in the relative sensitivity $\varepsilon(\lambda)$. Uncertainty of $1/\sqrt{N_{total}}$ in luminescence measurements due to shot noise was negligibly small. Thus, the uncertainty originated almost completely from the calibrations, and the total uncertainty was estimated to be 16%.

*Reagents and reaction.* The concentration and volume of reagents used for the measurements are listed in Table 3. They were basically the same as in the instructions of O'Kane and Lee for quantum-yield measurement of the aqueous luminol chemiluminescence standard (12, 13), though we checked various concentrations of solutions near that region. Luminol: A 1 x $10^{-4}$ M luminol (99% grade, Wako, Osaka, Japan) was characterized by absorbance with $\varepsilon_{347\ nm}$ = 7640 $M^{-1}cm^{-1}$. The luminol solution was diluted to 2 x $10^{-7}$ M – 1 x $10^{-5}$ M with buffer. 0.1 M potassium carbonate ($K_2CO_3$) solution was used as buffer to maintain pH 11.5. $H_2O_2$: 1.2 x $10^{-}$

$^{2}$ M, 0.12 M and 1.2 M aqueous solution of $H_2O_2$ (30%, Wako) were prepared and used. Horseradish peroxidase (HRP): HRP (Wako) in 0.1 M $K_2CO_3$ buffer solution was prepared and diluted to 4 x $10^{-7}$ M. The catalyst concentration was low, and the absorbance in the 420 nm region was below 0.02 $cm^{-1}$.

An aliquot (50 - 52 μl) of luminol and $H_2O_2$ solutions as listed in Table 3, in the acrylate tube cell shown in the setup of Figure 1 was added by the 50 μl HRP solution to initiate the reaction. The luminescence measurement was started just before the reaction was initiated and continued until the reaction was completed. All the measurements were done at the room temperature of 21 - 23 $^{o}$C.

RESULTS

**Luminescence spectra in absolute units**

Figure 5(a) shows the luminescence yield spectrum $N_\lambda$ ($nm^{-1}$) as a function of wavelength λ(nm) measured for the $M$ = 2.09 x $10^{14}$ luminol molecules in a mixed solution at pH 11.5 with 2 x $10^{-3}$ M $H_2O_2$ and 2 x $10^{-7}$ M HRP, which is an optimized condition as shown later. Note that the vertical axis expresses in the absolute number of photons as a function of wavelength instead of arbitrary units or RLU. Figure 5(b) shows the luminescence yield spectrum $N_E$ on the left vertical axis as a function of photon energy $E$(eV), derived by Eq. (3) from $N_\lambda$ in Figure 5(a).

Also shown in Figure 5(b) on the right vertical axis is the spectrum of quantum yield

$N_E$ / $M$ as a function of photon energy $E$(eV). Note that this shows the spectrum of quantum yield in the aqueous luminol chemiluminescence in the optimized condition, which shows not only the spectral shape of luminescence but also the quantum yield, or the luminescence efficiency. Since luminol is a widely used standard in chemiluminescence, this manner of presenting the luminescence spectrum is important for comparisons with other systems.

## Quantum yields

The integrated area of the spectra $N_\lambda$ and $N_E$ in Figure 5 gives the integrated total number of luminescence photons as $N_{\text{total}} = 2.57 \times 10^{12}$, and a quantum yield of 1.23% is obtained from $N_{\text{total}}$ / $M$. By definition, the same is obtained as the integrated area of the spectrum $N_E$ /$M$ shown on the right vertical axis in Figure 5(b).

Figure 6 plots the integrated numbers $N_{\text{total}}$ of luminescence photons versus numbers $M$ of luminol molecules, which were measured for solutions with various concentrations of luminol and $H_2O_2$. All the concentrations shown in the figure are concentrations after mixing of the respective reagent solutions to the final volumes of 100 – 102 μl, whereas Table 3 indicates initial concentrations of reagents. For luminol concentrations from $8 \times 10^{-8}$ M to $5 \times 10^{-6}$ M, or number of molecules from $5 \times 10^{12}$ to $3 \times 10^{14}$ in the mixed solutions, the integrated number of luminescence photons $N_{\text{total}}$ was proportional to the number of luminol molecules for all four different conditions of $H_2O_2$ ($1 \times 10^{-4}$ M, $5 \times 10^{-4}$ M, $2 \times 10^{-3}$ M, and $5 \times 10^{-3}$ M). Quantum yields were evaluated from the slopes of the fitting lines in the figure, where good linearity proves the validity of quantum-yield determination. The highest quantum yield of 1.23 ± 0.04% was obtained for $H_2O_2$ concentration of 2 x $10^{-3}$ M shown by open circles in Figure 6. Since the total uncertainty of the measurement system was 16% as mentioned in the preceding section, the optimized quantum yield was determined to be 1.23 ± 0.20%.

Note here that quantum yields of the aqueous luminol chemiluminescence changed with $H_2O_2$ concentration. As the $H_2O_2$ concentration was increased from 1 x $10^{-4}$ M to 2 x $10^{-3}$ M, the quantum yield increased from 0.8 ± 0.13% to the maximum value of 1.23 ± 0.20%. However, when $H_2O_2$ was further increased to 5 x $10^{-3}$ M, the quantum yield decreased to 1.0 ± 0.17%.

As discussed below, the optimized quantum yield of 1.23 ± 0.20% in the present experiment shows very good agreement with the value of 1.24% given by the majority of previous related reports (6-13).

**DISCUSSION**

The investigation of quantum yield in luminol chemiluminescence has a long history and the first thorough study and review was by Lee and Seliger in 1965 (6), who reported a quantum yield of 1.25 ± 0.06% for luminol chemiluminescence optimized in air saturated aqueous solution with hemin-catalyzed hydrogen peroxide at pH 11.6. The quantum yield of aqueous luminol chemiluminescence was later reexamined with independent elaborate methods (8-13), and most of them supported this result by reporting values of 1.1 - 1.2%. More recently, O'Kane and Lee (12,13) concluded that the absolute quantum yield of luminol chemiluminescence is 1.24% in both an HRP-catalyzed $H_2O_2$ aqueous solution and a dimethylsulfoxide solution, described their recommended experimental procedures, showed its usefulness as a low-level light standard, and listed previous reports on luminol chemiluminescence. The optimized quantum-yield result of 1.23 ± 0.20% in our present experiment is in good agreement with these previous reports. This indicates that the calibration method used here is consistent with most of the previous reports.

Luminol chemiluminescence is now often used as a secondary light standard. However, we recognized some concerns about this. First of all, we found that the quantum yield changed sensitively with the concentration of $H_2O_2$ and there was an optimum $H_2O_2$ concentration. Since $H_2O_2$ degrades by decomposition with time, it may not be easy to keep the optimum concentration for a long time. Optimization should be repeated occasionally.

Though good linearity between luminescence intensity and luminol concentration was confirmed in all conditions in our experiment, the linearity check is always important in characterizing the quantum yield. This is because each detection system has its own background signal level due to dark currents in the detector, background light, contamination of solution, and other sources, which may disturb the estimation of luminescence intensity.

In addition, though commercially available luminol powder is labeled as having 99% purity, this value may be misleading. Supplies say that the purity is determined by HPLC as a chromatogram area percentage but not as a mass percentage. Thus, the concentration of luminol solution must be determined by absorbance measurements. In fact, the concentration determined by our absorbance measurement reveals that the real purity of commercial luminol powder is about 90% w/w.

Due to these subtleties in luminol and its chemiluminescence, many additional supporting characterizations are necessary for it to be used as a secondary light standard. This example of luminol chemiluminescence indicates a strong need to establish a measurement standard of luminescence for any colors and commercial equipment based on it.

Quantitative measurements of luminescence and quantum yields have been intensively performed for bioluminescence such as firefly (1), cypridina (2), and aequorin (3-5), as well as for the luminol chemiluminescence standard (6-13) and the radioactive light standard of Hastings (14). In most previous experiments, however, light-collection-efficiency calibration similar to ours was not performed, but a point-source approximation for a whole solution was made. In other words, researchers assumed that all internal luminescence from substrate molecules came from the sample cell, and that it was isotropic, where the effects of total internal reflection and light piping can be neglected. To reduce estimation errors due to this assumption, some of them used a spherical vessel (2), an integration sphere (10), and/or a large distance between the sample cell and the detector. An important advantage of our light-collection-efficiency calibration over previous techniques is that it is applicable to samples cells of any kind, which allows flexibility for various samples, reaction conditions, and luminescence measurements.

As an absolute light power standard in the absolute sensitivity calibration, we used a laser and a calibrated optical power meter, while previous methods used standard lamps or actinometry. In principle, differences in calibration methods do not matter, if they are done correctly. In practice, however, simplicity and ease of operation are important to reduce the risks of errors and uncertainty. We wish to emphasize that directional light from a laser is a very convenient tool for calibrating detector sensitivity. Furthermore, characterization of laser light power by a commercial calibrated optical power meter is easy and reliable, which is like the measurement of electrical voltage by a tester or digital multimeter. In contrast, the use of a standard lamp needs careful complete shielding of stray light, which is like keeping a room dark with the light on.

In our measurement system we used a cooled CCD camera as a sensitive photodetector, while most of the previous experiments used photomultiplier tubes (PMTs). Though PMTs have advantages in cost and fast response, the cooled CCD camera has great advantages of high quantum efficiency, very low DC noises, and parallel multichannel photo-detection, which are very suitable for measuring bio/chemiluminescence spectra efficiently. Because of these advantages, efficient spectral measurements for small amounts of sample solution are possible, which allows downsizing of the whole system.

Once our present measurement system was built as tabletop equipment with long-term stability, calibration should be effective for some period, enabling luminescence yield spectrum measurement, quantum yield determination, and its spectroscopy in each biochemistry laboratory.

*Acknowledgments*--This work was financially supported partly by a Grant-in-Aid from MEXT, Japan.

## FIGURE LEGENDS

**Figure 1**. Experimental setup for absolute luminescence yield measurement.

**Figure 2**. The reference double-plate cell and sample cells for the luminescence measurement.

**Figure 3**. Schematic drawing of the reference double-plate cell and lens and iris system with numerical aperture *NA*.

**Figure 4.** Volume dependence of light-collection efficiencies for four kinds of cells: acrylate tube cell and white, clear, and black 384-well microplate cells.

**Figure 5**. a) Luminescence yield spectrum in the unit of $nm^{-1}$. b) Luminescence yield spectrum with left axis in units of $eV^{-1}$ and quantum yield spectrum with right axis in units of $eV^{-1}$ measured for $2.09 \times 10^{14}$ luminol molecules at pH 11.5.

**Figure 6**. Luminescence yield with changing luminol molecule number for some different conditions of $H_2O_2$ ($1 \times 10^{-4}$ M, $5 \times 10^{-4}$ M, $2 \times 10^{-3}$ M, and $5 \times 10^{-3}$ M in mixed solutions). QY: quantum yield.

Table 1. Light-collection efficiency $\eta_{cell}$(%) for several kinds of cells with 100 μl of luminescent solution.

| Cell type | Color | Ratio | $\eta_{cell}$(%) |
|---|---|---|---|
| Reference double-plate cell | transparent | 1 | 0.262 (cal.) |
| Acrylate tube cell | transparent | 1.0 | 0.26 ± 0.01 |
| 384-well microplate | transparent | 0.7 | 0.18 ± 0.02 |
| | black | 0.6 | 0.16 ± 0.02 |
| | white | 6.5 | 1.7 ± 0.2 |

Table 2. Sources of uncertainty.

| Source | Uncertainty(%) | Total(%) |
|---|---|---|
| Collection efficiency, $\eta_{cell}\,T$ | | 10.6 |
| $\eta_{plate}$ via $NA$ (geometry) | 0.3 x 2 | |
| relative efficiency, $\eta_{cell}$ / $\eta_{plate}$ | 10 | |
| slit transmissivity, $T$ | 3.4 | |
| CCD-spectrometer sensitivity, $S(\lambda)$ | | 10.9 |
| absolute sensitivity, $S_{\lambda 0}$ | 9.4 | |
| relative sensivity, $\varepsilon(\lambda)$ | 7.1 | |
| CCD counts for luminescence, $C(\lambda)$ | | <<1 |
| Total uncertainty (±%) | | 16 |

Table 3. Concentrations and volumes of reagents.

| Reagents | Concentration (M) | Volume (μl) |
|---|---|---|
| Luminol (M.W.: 177.16) | $2 \times 10^{-7} – 1 \times 10^{-5}$ | 50 |
| $H_2O_2$ | $1.2 \times 10^{-2} – 1.2$ | 0.5 - 2 |
| HRP (M.W.: 44.000) | $4 \times 10^{-7}$ | 50 |

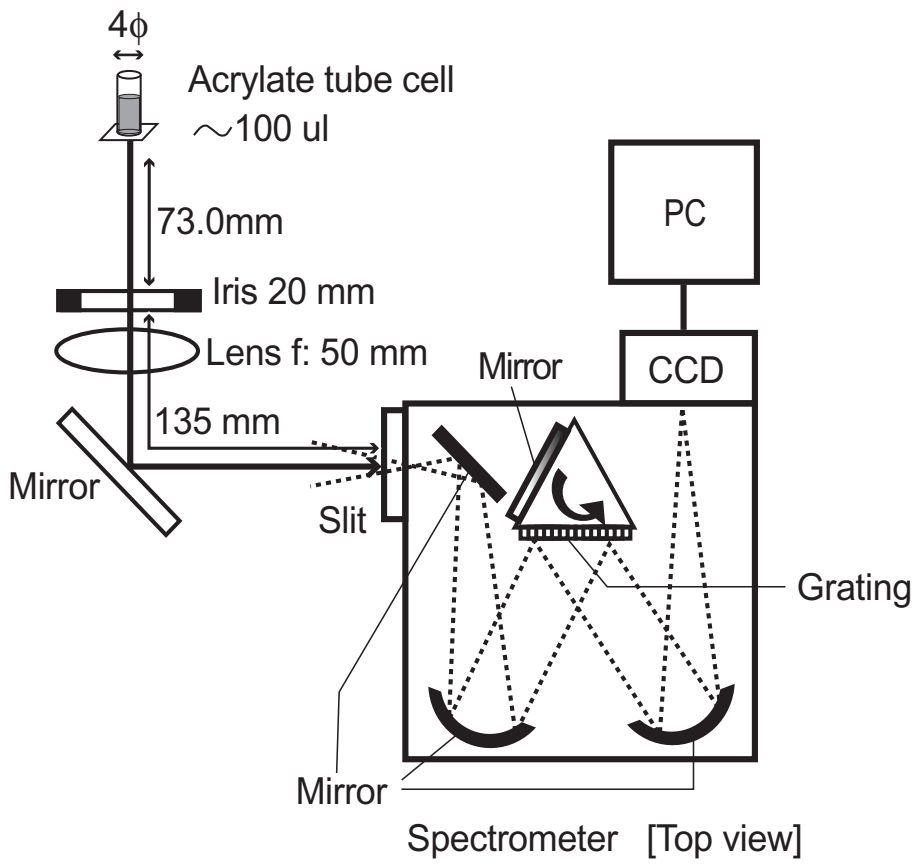

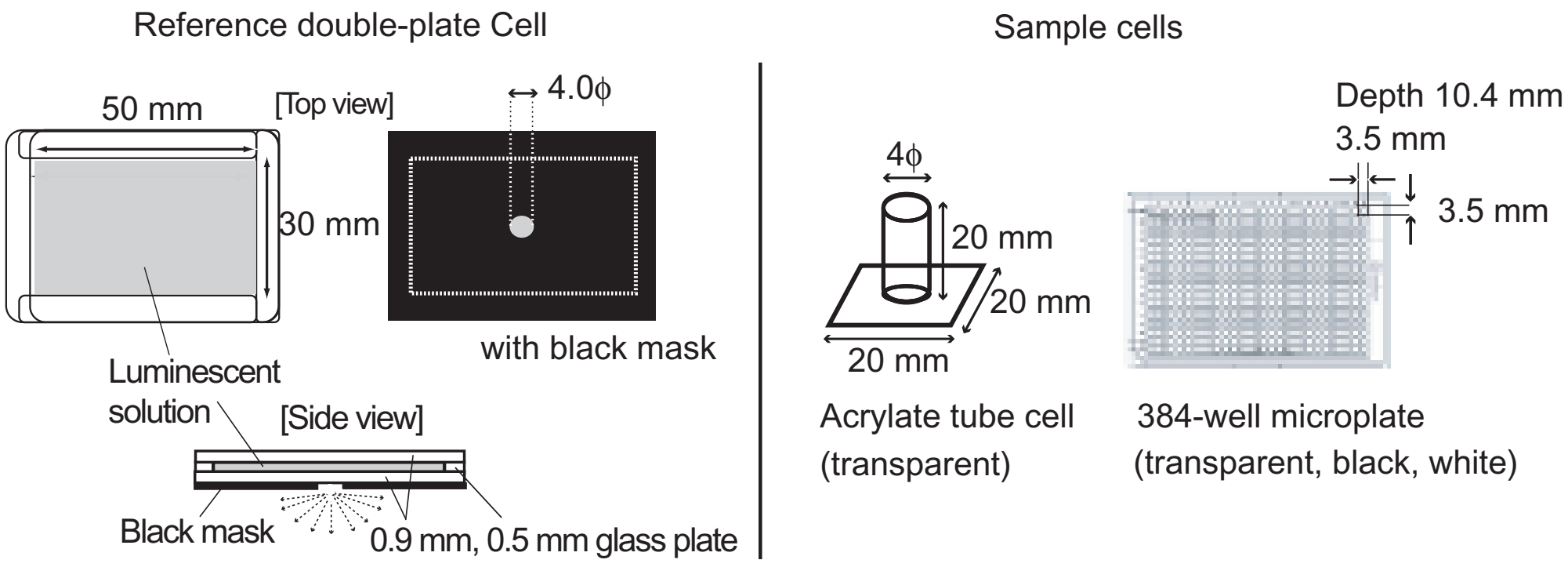

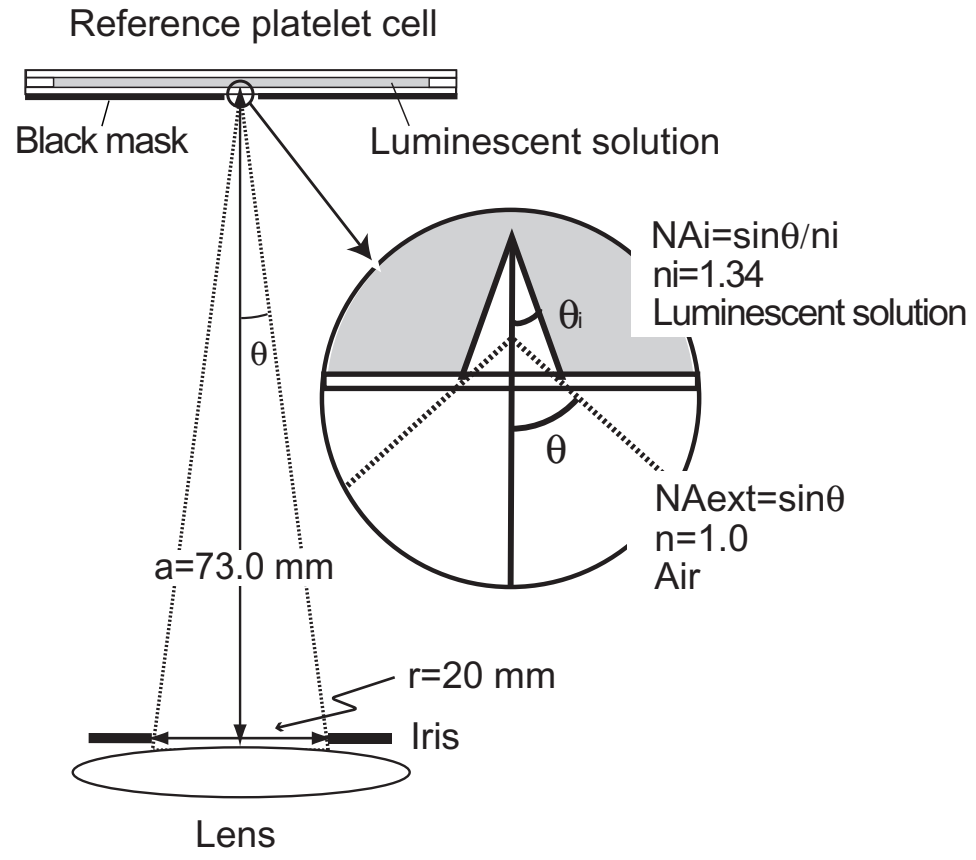

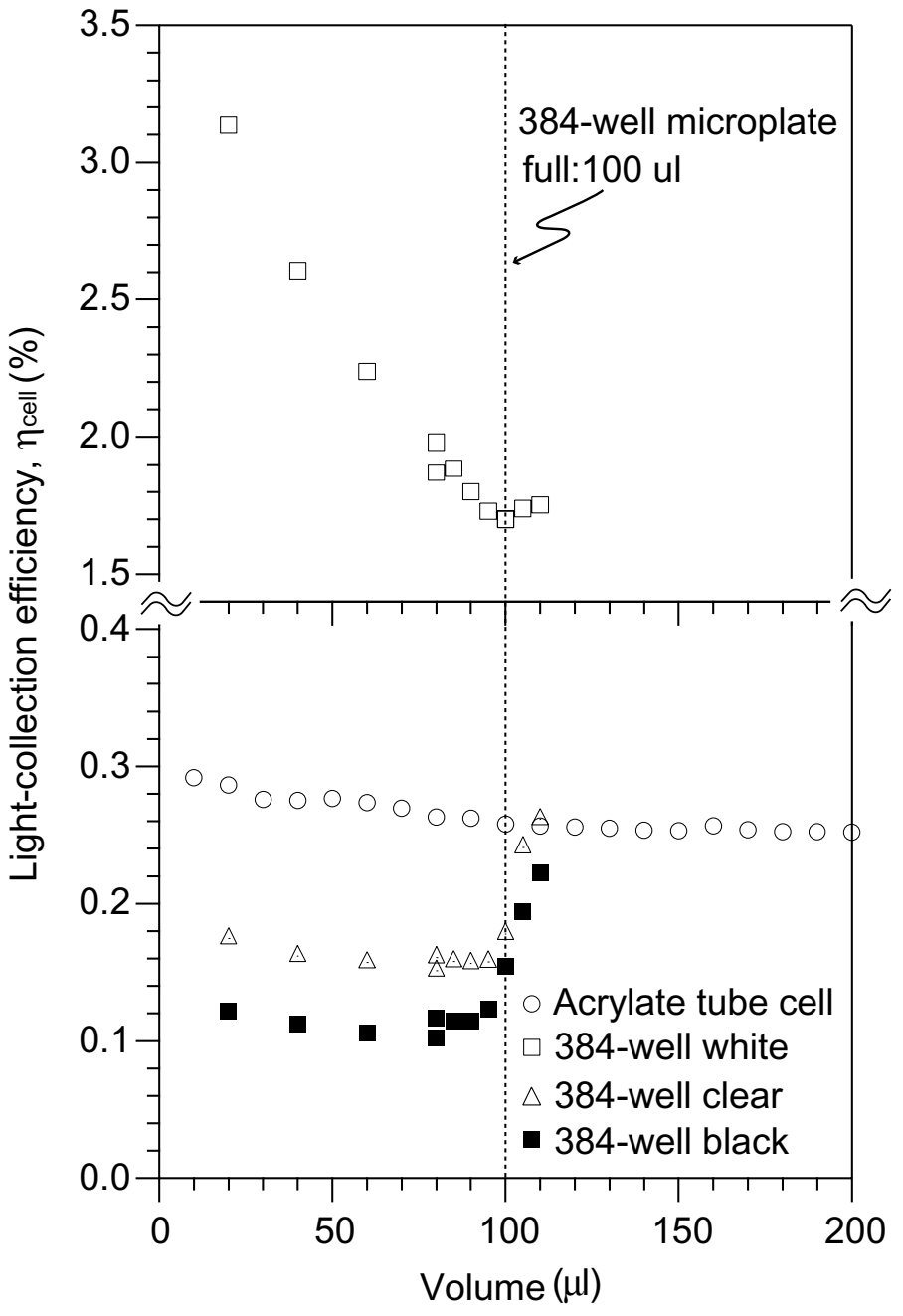

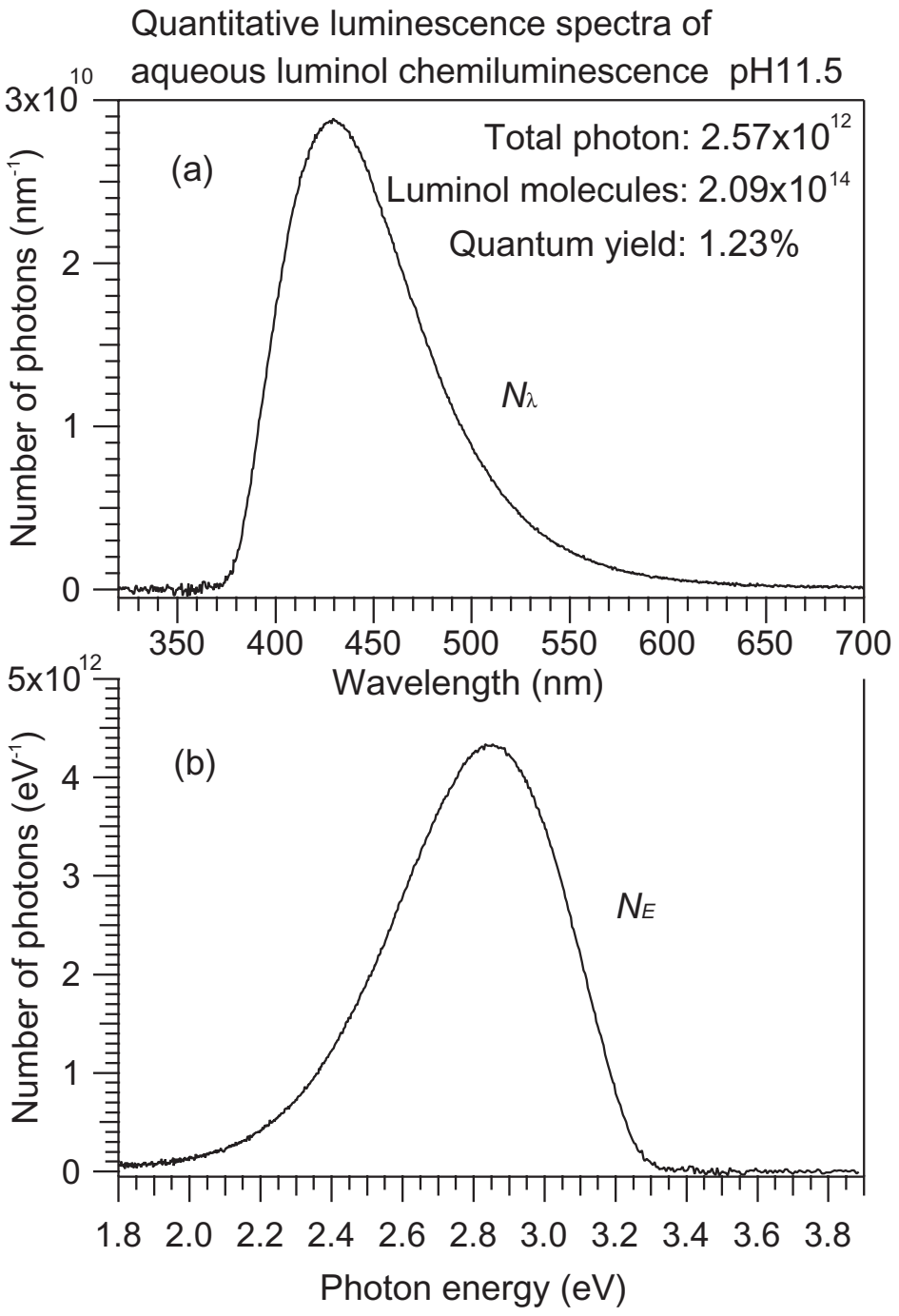

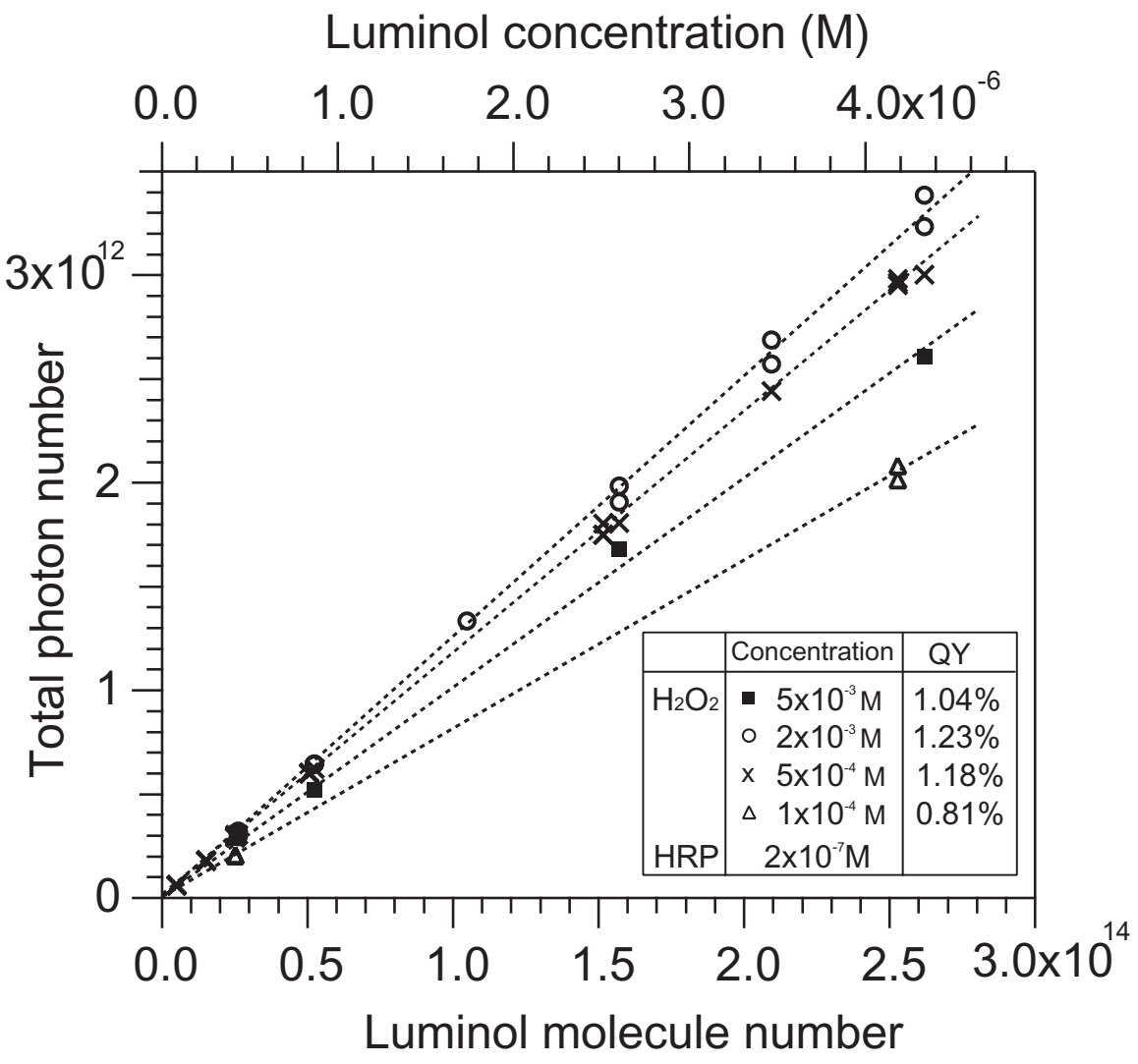